\documentclass[sigconf,screen]{acmart}
\acmConference[FSE 2025]{ACM ACM International Conference on the Foundations of Software Engineering}{23-27 June, 2025}{Trondheim, Norway}
\usepackage{graphicx}
\usepackage{color}
\usepackage{xspace}
\usepackage{amsmath,amsfonts}
\usepackage{algorithm}
\usepackage{algorithmic}
\usepackage{textcomp}
\usepackage{xcolor}
\usepackage{booktabs}
\usepackage{booktabs,xltabular}
\usepackage{wrapfig}
\usepackage{float} % for fixing the position of minipage table
\usepackage{caption} % for adding captions to minipages

\usepackage{hyperref}
\usepackage{hyperxmp}

\newcommand{\squishlist}{
 \begin{list}{$\bullet$}
   { \setlength{\itemsep}{0pt}
     \setlength{\parsep}{0pt}
     \setlength{\topsep}{0pt}
     \setlength{\partopsep}{0pt}
     \setlength{\leftmargin}{2.5em}
     \setlength{\labelwidth}{1.5em}
     \setlength{\labelsep}{0.5em} } }

\def\BibTeX{{\rm B\kern-.05em{\sc i\kern-.025em b}\kern-.08em
    T\kern-.1667em\lower.7ex\hbox{E}\kern-.125emX}}

\newcommand{\squishend}{
  \end{list}  }

\newcommand{\TheName}{\textsc{HumanLLM}}

\acmBooktitle{Companion Proceedings of the 33rd ACM Symposium on the Foundations of Software Engineering (FSE '25), June 23--27, 2025, Trondheim, Norway}

\setcopyright{rightsretained}

\begin{document}

\acmYear{2025}\copyrightyear{2025}
\acmConference[FSE '25]{33rd ACM International Conference on the Foundations of Software Engineering}{June 23--28, 2025}{Trondheim, Norway}
\acmBooktitle{33rd ACM International Conference on the Foundations of Software Engineering (FSE '25), June 23--28, 2025, Trondheim, Norway}
\acmDOI{10.1145/3696630.3728510}
\acmISBN{979-8-4007-1276-0/25/06}

\title{Enhancing Code LLM Training with Programmer Attention}

% \author{Anonymous Author(s)}

\author{Yifan Zhang}
\orcid{0000-0001-5719-772X}
\affiliation{%
  \institution{Vanderbilt University}
  % \streetaddress{123 Example Street}
  \city{Nashville}
  % \state{TN}
  % \postcode{12345}
  \country{USA}}
% \email{alice@example.com}

\author{Chen Huang}
\orcid{0000-0002-3542-7085}
\affiliation{%
  \institution{Sichuan University}
  % \streetaddress{456 Research Blvd.}
  \city{Chengdu}
  % \state{Sichuan}
  \country{China}}
% \email{bob.johnson@example.com}

\author{Zachary Karas}
\orcid{0000-0002-5721-8794}
\affiliation{%
  \institution{Vanderbilt University}
  \city{Nashville}
  \country{USA}}

\author{Dung Thuy Nguyen}
\orcid{0000-0002-3489-4255}
\affiliation{%
  \institution{Vanderbilt University}
  \city{Nashville}
  \country{USA}}

\author{Kevin Leach}
\orcid{0000-0002-4001-3442}
\affiliation{%
  \institution{Vanderbilt University}
  \city{Nashville}
  \country{USA}}

\author{Yu Huang}
\orcid{0000-0003-2730-5077}
\affiliation{%
  \institution{Vanderbilt University}
  \city{Nashville}
  \country{USA}}

\begin{abstract}
Human attention provides valuable yet underexploited signals for code LLM training, offering a perspective beyond purely machine-driven attention. Despite the complexity and cost of collecting eye-tracking data, there has also been limited progress in systematically \emph{using} these signals for code llm training. To address both issues, we propose a cohesive pipeline spanning \emph{augmentation} and \emph{reward-based fine-tuning}. Specifically, we introduce (1) an \emph{eye-tracking path augmentation} method to expand programmer attention datasets, (2) a \emph{pattern abstraction} step that refines raw fixations into learnable attention motifs, and (3) a \emph{reward-guided} strategy for integrating these insights directly into a CodeT5 supervised fine-tuning process. Our experiments yield +7.16 in CodeBLEU on the CodeXGlue benchmark for code summarization, underscoring how uniting human and machine attention can boost code intelligence. We hope this work encourages broader exploration of human-centric methods in next-generation AI4SE. 
\end{abstract}




\maketitle

\section{Introduction}
\label{sec:intro}

Programmers often exhibit selective attention when interpreting and modifying source code. Understanding these real-world attention patterns can deepen our grasp of code comprehension, improving tasks like summarization or completion~\cite{hoq2024towards,nijkamp2023codegen2}. However, collecting such data (for example, via eye tracking or manual annotations) is expensive and time-consuming, limiting its large-scale use. Even when available, novel ways of integrating human attention signals into LLM training remain limited~\cite{chen2021evaluating,kou2024large}, so most AI-driven code tools rely on static artifacts alone. This gap underscores the need for data-augmentation strategies and specialized pipelines that systematically embed genuine programmer fixations into LLM-based code intelligence.

While LLMs like CodeT5 improve when fine-tuned on large code corpora, integrating human signals like eye-tracking attention patterns can enhance them further~\cite{zhang2024eyetrans,kou2024large}. However, typical data augmentation often overlooks code's cognitive demands or specific structures~\cite{hamza2024human,chen2021evaluating}. Consequently, even advanced LLMs may underutilize valuable programmer attention information (like scanpaths or reading orders), especially when such data is sparse. Effectively embedding these developer signals necessitates specialized training pipelines. To address this, we propose \TheName{}, a two-part approach primarily for code summarization (also tested on completion and translation). First, \TheName{} uses real eye-tracking data reflecting programmer reading to augment code. Second, it embeds these signals via reward-based fine-tuning of CodeT5~\cite{wang2021codet5}, aligning the model with human fixations. This systematic use of human-centric patterns aims to bridge cognitive insights and automated code intelligence across tasks.

Although our attention patterns originate from code summarization studies and may not fully generalize to completion or translation, our experiments show that injecting these human-centric signals can notably improve model performance. For instance, we observed a up to +7.16 in CodeBLEU~\cite{ren2020codebleu}, +2.67 in Syntax, and +14.97 in Dataflow on the CodeXGlue dataset for code summarization, suggesting the potential of attention-driven methods to enhance software engineering workflows. This reinforces the vision that systematically integrating real programmer attention into code LLM training can unify cognitive insights with automated solutions across diverse AI4SE tasks.

\section{Preliminaries}\label{sec:preliminaries}

Our two-part code LLM fine-tuning pipeline leverages three key preliminaries to integrate developer attention:

\paragraph{Eye-Tracking for Programming.}
Eye tracking captures \emph{fixations}, where developers pause on tokens, and 
\emph{saccades}, the rapid jumps between them, revealing real-time cognitive 
strategies in code reading~\cite{minelli2015know,salvucci2000identifying}. 
These insights show how programmers parse syntax and identify relevant 
constructs, grounding our human-centered pipeline (Section~\ref{sec:approach}).

\paragraph{Data Augmentation in AI for Code.}
Traditional augmentation methods (e.g., token substitutions, AST edits) often 
overlook developers’ natural reading flows~\cite{allamanis2018survey,yu2022data}. 
By integrating actual eye-tracking data (Section~\ref{sec:method-augmentation}), 
our approach captures these real patterns, helping models learn cognitively 
grounded cues that reflect authentic developer attention.

\paragraph{Code LLM Training.}
Modern code LLMs typically follow \emph{Retrieval-Augmented Generation (RAG)} 
or \emph{Supervised Fine-Tuning (SFT)}~\cite{chen2021evaluating,he2024exploring}. 
We employ a \emph{reward-guided SFT} procedure, periodically injecting 
eye-tracking signals into the training loop (Section~\ref{sec:finetuning}), so 
that model outputs better match actual programmer fixations.

\section{Approach}\label{sec:approach}

\begin{figure}[tb]
    \centering
    \includegraphics[width=\columnwidth]{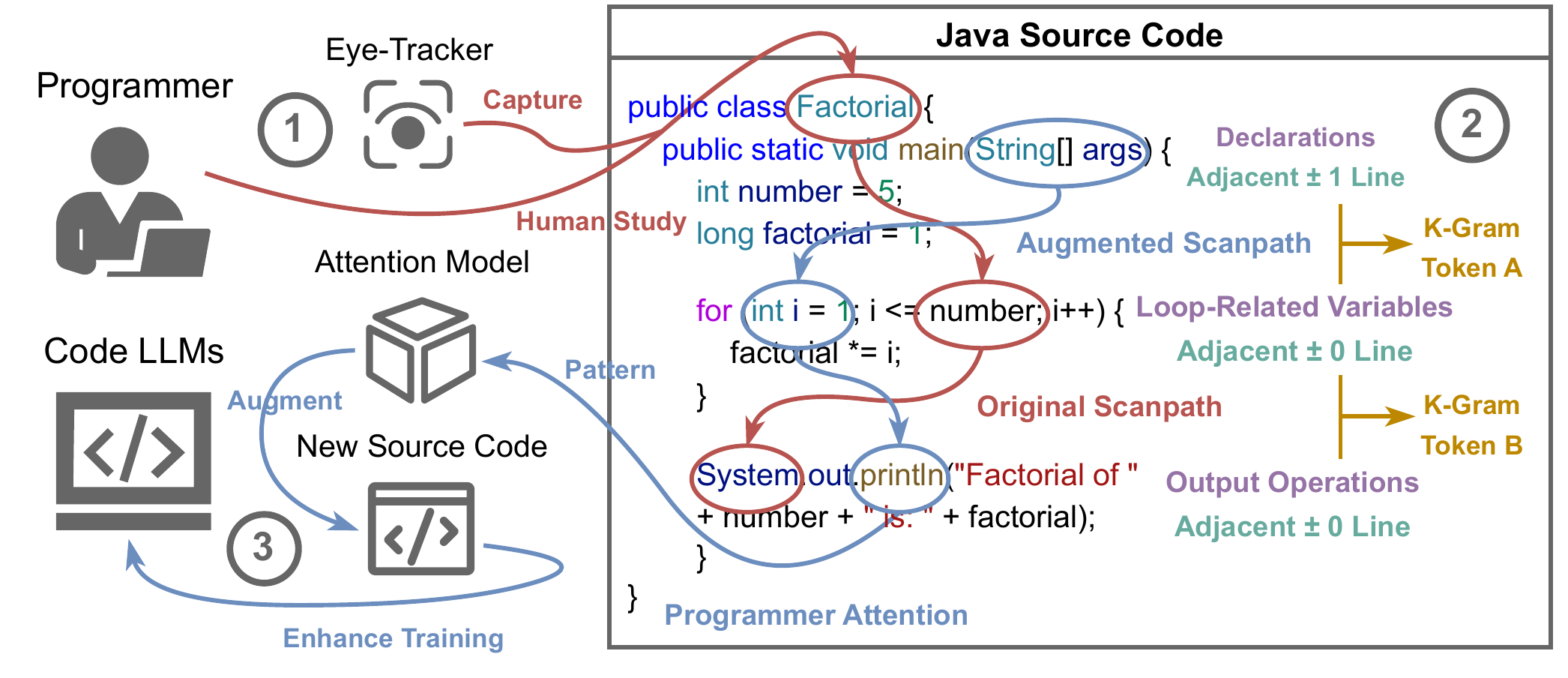}
    \caption{Overview of our pipeline~(\TheName{}): \textcircled{1} collects eye-tracking data 
    (red) to capture real programmer attention; \textcircled{2} augments these fixations 
    with AST-based adjacency (blue) and k-gram patterns; and \textcircled{3} 
    uses these human signals to guide a reward-based CodeT5 fine-tuning.}
    \label{fig:augmented_scanpath}
\end{figure}

Figure~\ref{fig:augmented_scanpath} outlines our pipeline, which we refer to as \TheName{}, integrating eye-tracking data into a CodeT5 model in three stages. First, \textcircled{1}~\emph{Data Collection} gathers token-level fixation data from an open Java eye-tracking dataset~\cite{karas2024tale} and supplements it with Java snippets from CodeXGlue\footnote{\url{https://github.com/microsoft/CodeXGLUE}}. Second, \textcircled{2}~\emph{Augmentation} enriches these fixations with AST-based semantic labels, adjacency expansions, k-gram patterns, and positional ordering (Sections~\ref{sec:datacollection}--\ref{sec:method-augmentation}). Finally, \textcircled{3}~\emph{Reward-Based Fine-Tuning} periodically fuses these human-guided signals into CodeT5 training (Section~\ref{sec:finetuning}), aligning the model with genuine programmer behavior.

\subsection{Data Collection}
\label{sec:datacollection}

We use the open eye-tracking dataset~\cite{karas2024tale}, which captures how 15 professional developers read and summarized 120 Java methods (30–50 lines each), logging token-level fixations and saccades. To enhance diversity, we incorporate Java samples from CodeXGlue. Each snippet provides \emph{code tokens} (e.g., \texttt{if}, \texttt{UserInput}, \texttt{42}), mapped to \emph{semantic tokens} (\texttt{Keyword}, \texttt{Identifier}, \texttt{Literal}) via AST annotations. These fixation traces and mappings serve as the foundation for augmentation and reward-based training.

\subsection{Human-Centric Augmentation of Code Tokens}
\label{sec:method-augmentation}

Let \(\mathbf{T} = \{t_1,\dots,t_n\}\) be the tokens in a Java snippet, and let \(\mathbf{F}\subset\mathbf{T}\) represent those actually fixated by developers. We construct an enriched set \(\mathbf{F}^\star\) that encodes the following programmer attention information:

\paragraph{Semantic Labeling.}
We parse each method’s AST and assign a label \(L(t_i)\) (e.g., \(\text{Keyword}\), \(\text{Literal}\)) to each token \(t_i\in\mathbf{T}\). This step ensures that the model recognizes syntactic roles rather than relying on raw text.

\paragraph{Adjacency-Based Expansion.}
For each \(t_i\in\mathbf{F}\), we include any tokens \(\pm3\) lines away that share \(L(t_i)\). Formally,
\[
    \mathbf{F}^\star \;\leftarrow\; \mathbf{F}\;\cup\;
    \bigl\{\,t_j\,\big|\,
    | \text{line}(t_j)-\text{line}(t_i)|\le3\;\wedge\;L(t_j)=L(t_i)\bigr\}.
\]
This threshold of three lines stems from our analysis of the open eye-tracking 
dataset~\cite{karas2024tale}, showing that over 95\% of attention shifts occur 
within this distance, indicating that developers often check nearby lines with 
similarly categorized tokens.

\paragraph{K-Gram Patterns.}
Prior work on developer behavior~\cite{karas2024tale,zhang2024eyetrans} shows that bigrams (\(k=2\)) and trigrams (\(k=3\)) reflect how programmers cluster code tokens. We detect these patterns in \(\mathbf{F}^\star\), retain the top 20 for reward modeling, and assign each \(p_\ell\) a numeric label. If a snippet matches both, we prioritize the trigram, breaking ties by frequency. Capturing these reading clusters models how developers group tokens during comprehension.

\paragraph{Positional Ordering.}

Beyond adjacency expansions and k-gram patterns, we track each token’s reading index \(\pi(t_i)\) to align augmented tokens with actual scanpaths. This captures how developers encounter tokens in sequence, extending beyond simple locality. We record positions for up to 100 tokens per snippet, ensuring consistent fixation mapping. Each token \(t_i \in \mathbf{F}^\star\) is labeled with \(\bigl(p_\ell(t_i), \pi(t_i)\bigr)\), encoding its semantic pattern and reading order for a more human-guided code representation.

\subsection{Reward-Based Fine-Tuning with Programmer Attention}
\label{sec:finetuning}

We integrate these augmented signals into a CodeT5 pipeline that periodically references the human-informed dataset \(\mathbf{F}^\star\). After every \(m\) mini-batches~(\(m=20\)) of standard cross-entropy (CE) training, we sample a mini-batch from \(\mathbf{F}^\star\), generate predictions \(\hat{y}\), and compare them to the ground-truth labels \(y^\star\). The resulting reward \(\mathcal{R}(\hat{y}, y^\star)\) measures how well the model’s predicted sequences align with genuine developer fixations (e.g., matching k-grams or positional order). We then form a combined loss
\[
    \mathcal{L}_{\mathrm{total}} \;=\;
    \mathcal{L}_{\mathrm{CE}} \;+\;\alpha\,\mathcal{R},
\]
where \(\alpha\) regulates the strength of human guidance. By repeatedly backpropagating \(\nabla \mathcal{L}_{\mathrm{total}}\), CodeT5 learns to minimize typical prediction error while conforming to real programmer attention patterns. Over multiple epochs, these reward passes steer the model to internalize adjacency relationships, k-gram fixations, and realistic scanpath ordering, thereby producing more human-aligned outputs (e.g., code summaries).

\section{Experimental Design}\label{sec:experimental_design}

We conduct two sets of experiments: 
(1) a basic Transformer analysis on an 80--20 split of our augmented eye-tracking dataset to assess whether the enriched signals are genuinely learnable, and 
(2) a CodeT5-based evaluation (using \texttt{codet5-base}\footnote{\url{https://github.com/salesforce/CodeT5}}) on three tasks in CodeXGlue for Java (summarization, completion, and Java-to-C\# translation). 

In the first setting, we train for a single epoch, tracking progress (0--100\%) via 
the \emph{batch-size ratio} to ensure the Transformer absorbs semantic~(k-gram) and positional signals. In the second, we fine-tune CodeT5 with a reward-based loop 
(every 20 batches), adding an \(\mathcal{R}\)-driven term to cross-entropy that 
aligns predictions with real developer fixations. We fix seed 42 for data 
sampling and initialization, train on two NVIDIA A6000 GPUs (48\,GB each), and 
use AdamW (\(\text{lr}=5\times10^{-5}\)) with gradient checkpointing.

To balance diversity and runtime constraints, we sample 10\% of each CodeXGlue dataset. We report BLEU, ROUGE-L, and METEOR for textual fidelity in code summarization, and CodeBLEU for deeper code-level accuracy (incorporating syntax and dataflow checks). This combination of metrics provides a comprehensive measure of both linguistic coherence and program correctness. 

\section{Experimental Results}\label{sec:experimental_results}

We structure our analysis in three parts: (1) validating augmented signals across different adjacency windows and batch-size schedules, (2) showing how these signals improve CodeT5’s summarization (both NLP and code metrics), and (3) testing their transfer to completion and translation.

\subsection{Learning with Human-Centric Augmentations}

\begin{figure}[tb]
    \centering
    \includegraphics[width=\columnwidth]{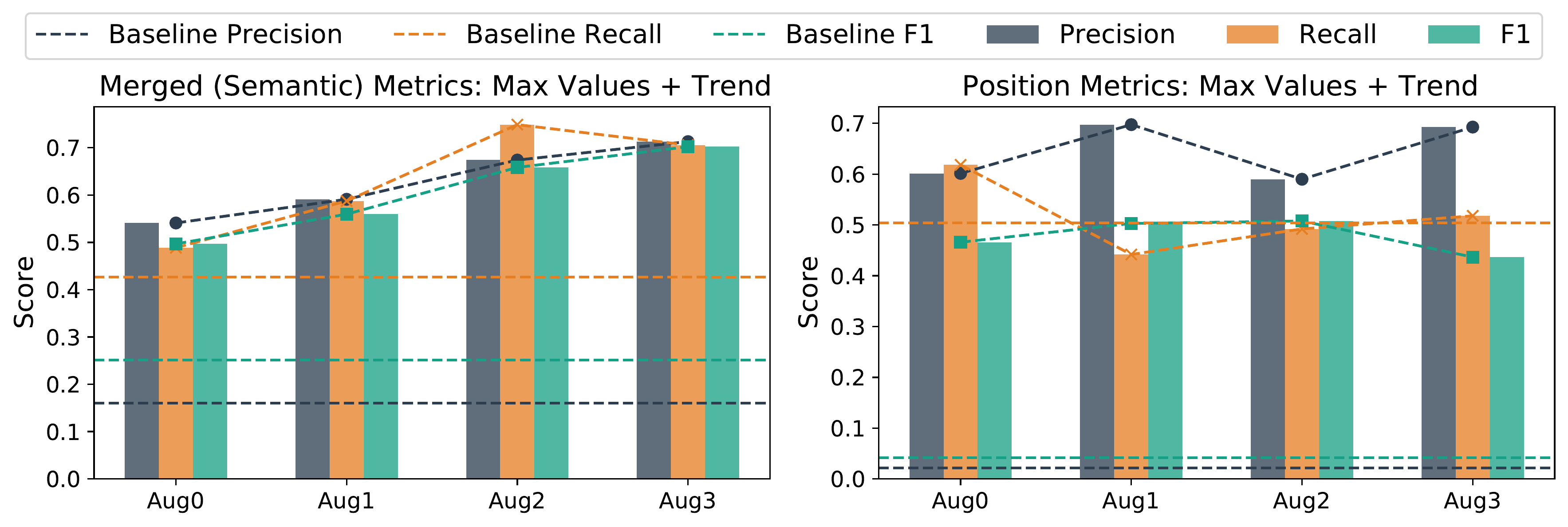}
    \caption{Impact of adjacency expansions (0 to 3 lines) on semantic (left) and positional (right) labels. 
    Wider windows generally improve Precision, Recall, and F1, exceeding a baseline Transformer (dashed lines) in five of six metrics.}
    \label{fig:augwindows}
\end{figure}

\paragraph{Adjacency Window Effects.}
Figure~\ref{fig:augwindows} reports Precision, Recall, and F1 on a held-out test set for semantic and positional labels when expanding adjacency from 0 to 3 lines. In five of the six metrics, performance surpasses a baseline Transformer (dashed lines). Only Recall for positional labels lags slightly at the widest window, indicating that too much context can dilute the reading-order signal.

\begin{figure}[tb]
    \centering
    \includegraphics[width=\columnwidth]{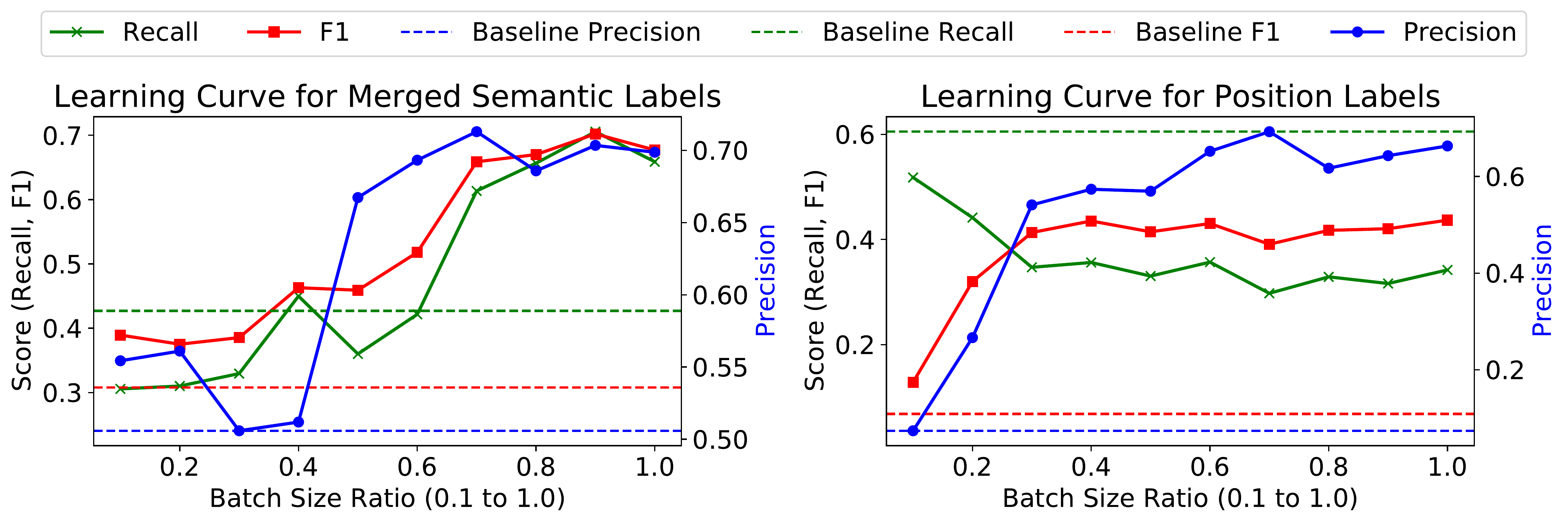}
    \caption{Learning curves for semantic (left) and positional (right) labels over one epoch, shown via the batch-size ratio. Dashed lines represent the baseline Transformer. Larger ratios correlate with better test-set performance.}
    \label{fig:batch_size_experiment}
\end{figure}

\paragraph{Batch-Size Ratio (Training Progress).}
Figure~\ref{fig:batch_size_experiment} then examines how performance evolves during a single epoch, where the \emph{batch-size ratio} on the horizontal axis reflects progress from 0\% to 100\% of that epoch. Larger ratios generally yield higher F1, Precision, and most Recall values, suggesting that the model benefits from encountering more human-centric samples earlier and more frequently.

Together, Figures~\ref{fig:augwindows} and~\ref{fig:batch_size_experiment} confirm that adjacency expansions and sufficient training exposure help the Transformer internalize real programmer fixations.

\subsection{Enhanced Code Summarization}
We next assess whether these cues boost code summarization. Table~\ref{tab:summarization-nlp} shows notable gains in BLEU, ROUGE-L, and METEOR, while Table~\ref{tab:summarization-code} reveals parallel improvements in CodeBLEU, Syntax, and Dataflow (up by nearly 15 points). This suggests that real developer fixations help the model track variables and data dependencies, mirroring human comprehension. 

\begin{table}[htbp]
    \centering
    \small
    \caption{\textbf{Summarization (NLP Metrics).} 
    \TheName{} integrates programmer attention into CodeT5, resulting in more fluent, context-aware summaries.}
    \label{tab:summarization-nlp}
    \scalebox{1.05}{
    \begin{tabular}{l | r | r | r}
        \toprule
        \textbf{Pipeline} & \textbf{BLEU} & \textbf{ROUGE-L} & \textbf{METEOR} \\
        \midrule
        Baseline  & 5.36  & 10.80 & 7.22 \\
        \TheName{}  & 10.32 & 14.91 & 12.88 \\
        \midrule
        Change    & +4.96 & +4.11 & +5.66 \\
        \bottomrule
    \end{tabular}
    }
\end{table}

\begin{table}[htbp]
    \centering
    \small
    \caption{\textbf{Summarization (Code Metrics).} 
    By leveraging actual fixations, \TheName{} demonstrates improved structural understanding and variable handling.}
    \label{tab:summarization-code}
    \scalebox{1.1}{
    \begin{tabular}{l | r | r | r}
        \toprule
        \textbf{Pipeline} & \textbf{CodeBLEU} & \textbf{Syntax} & \textbf{Dataflow} \\
        \midrule
        Baseline  & 8.51  & 3.06  & 18.01 \\
        \TheName{}  & 15.67 & 5.73  & 32.98 \\
        \midrule
        Change    & +7.16 & +2.67 & +14.97 \\
        \bottomrule
    \end{tabular}
    }
\end{table}

\begin{figure}[tp]
    \centering
    \includegraphics[width=\columnwidth]{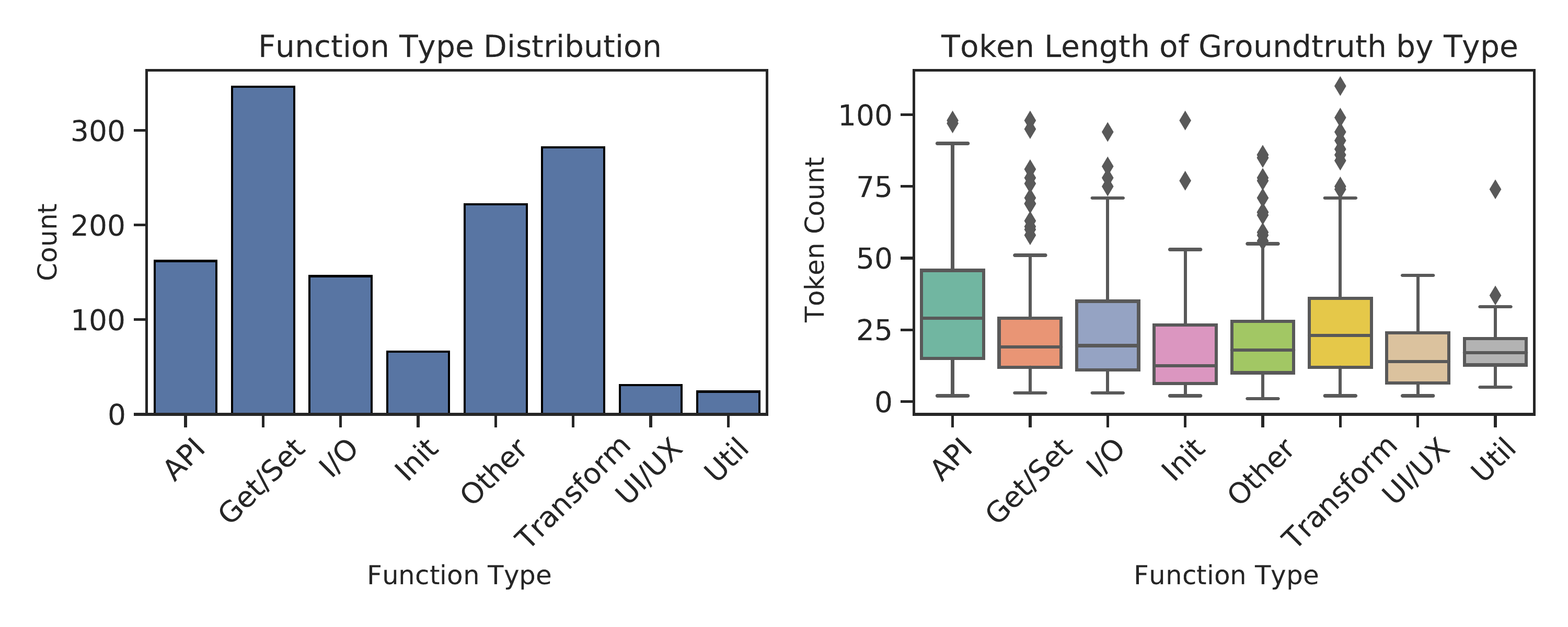}
    \caption{Overview of function types (left) and token-length distribution (right) for the ground-truth snippets in our dataset. The histogram on the left shows how frequently each function type appears, while the box plot on the right illustrates variability in token counts across different types.}
    \label{fig:function_type_distribution}
\end{figure}

\begin{figure}[tp]
    \centering
    \includegraphics[width=0.47\textwidth]{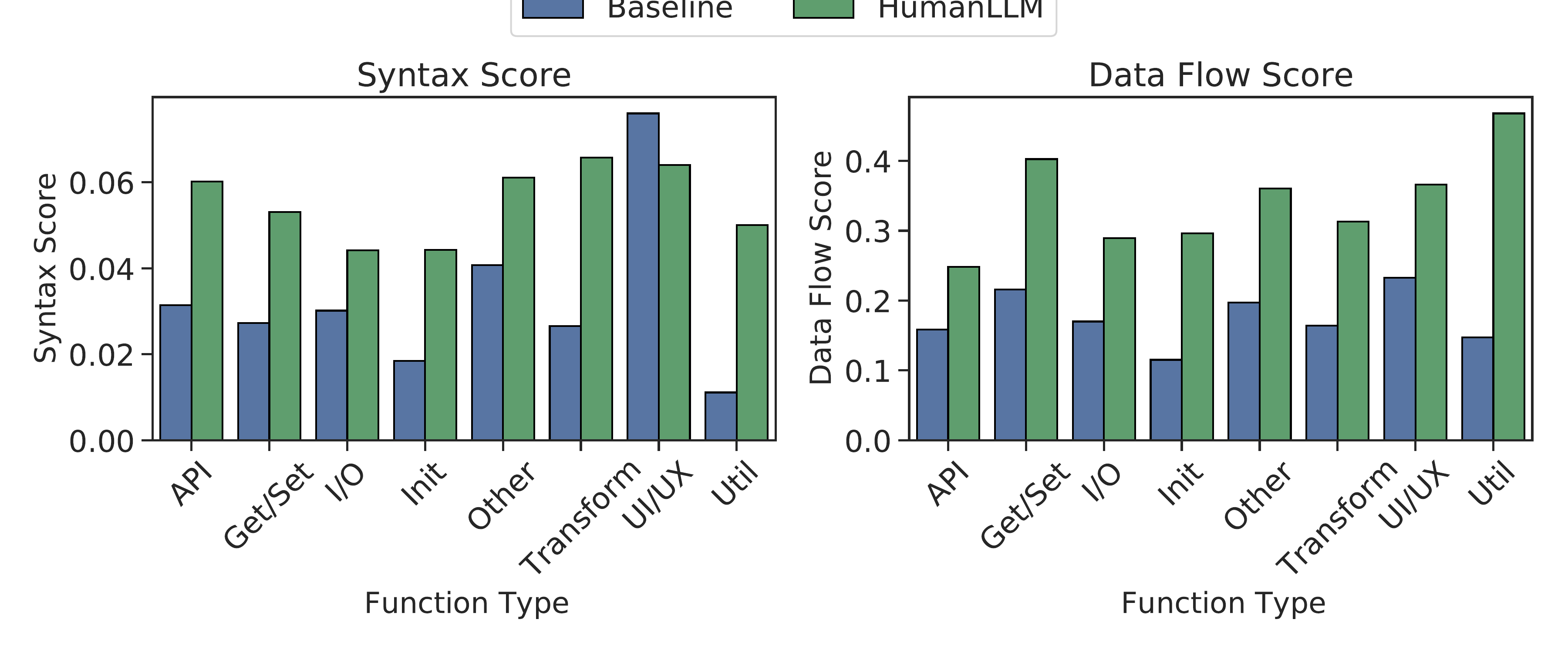}
    \caption{Syntax (left) and Data Flow (right) scores by function type for Baseline vs.\ \TheName{}. \TheName{} generally outperforms the Baseline by better capturing control structures and variable interactions, aligning more closely with real developer fixations.}
    \label{fig:syntax_dataflow}
\end{figure}

Figures~\ref{fig:function_type_distribution} and~\ref{fig:syntax_dataflow} show that these improvements extend across different function types. \TheName{} outperforms the baseline in syntax and data flow for most categories, indicating the augmented signals help CodeT5 adapt to varied coding styles.

\subsection{Task-Specificity for Completion and Translation}
Table~\ref{tab:completion-translation-code} shows results on code completion and Java-to-C\# translation, where syntax alignment sees minor gains but CodeBLEU and dataflow do not improve substantially. This highlights the specialization of our method for summarization: these fixation patterns do not transfer easily to tasks that rely on different developer attention cues. Future extensions may require distinct eye-tracking protocols or reward definitions to optimize performance in completion, translation, or other AI4SE scenarios.

\begin{table}[htbp]
    \centering
    \small
    \caption{\textbf{Completion and Translation (Code Metrics).} 
    Summarization-centric attention signals offer limited gains for other tasks.}
    \label{tab:completion-translation-code}
    \scalebox{1.0}{
    \begin{tabular}{l | l | r | r | r}
        \toprule
        \textbf{Task} & \textbf{Pipeline} & \textbf{CodeBLEU} & \textbf{Syntax} & \textbf{Dataflow} \\
        \midrule
        Completion
            & Baseline & 5.42 & 10.15 & 3.40 \\
            & \TheName{} & 4.77 & 10.46 & 1.45 \\
            \cmidrule(lr){2-5}   % Partial midrule for columns 2--5
            & Change   & -0.65 & +0.31 & -1.95 \\
        \midrule
       Translation
            & Baseline & 15.18 & 19.38 & 14.50 \\
            & \TheName{} & 14.94 & 18.96 & 14.38 \\
            \cmidrule(lr){2-5}   % Partial midrule for columns 2--5
            & Change   & -0.24 & -0.42 & -0.12 \\
        \bottomrule
    \end{tabular}
    }
\end{table}

\section{Future Work}\label{sec:future_work}

Our observations using eye tracking data from professional developers show promising improvements in code summarization metrics. In future work, we will develop novel augmentation techniques to better leverage these signals and extend benefits to tasks such as code review, fault localization, and security auditing. By incorporating richer eye tracking data and flexible reward strategies, we aim to build AI driven coding assistants that capture developer focus and deepen code understanding.

\section{Threat to Validity}\label{sec:threat_to_validity}

Our study is based on an eye tracking dataset from professional Java developers, which may limit the generalizability of our findings to other languages and paradigms. The use of 10\% of the CodeXGlue data may introduce sampling biases and, although we fix the random seed (42), some variation remains. Our reward scheme is designed for code summarization and may require adjustments for tasks such as bug detection, refactoring, or translation. Finally, the resource intensive nature of gathering eye tracking data calls for more efficient methods to leverage these signals.

\section{Related Work}\label{sec:related_work}
We briefly survey how code intelligence integrates human-centered signals, review neural approaches to code comprehension, and note why many LLMs overlook real programmer focus. Our method closes these gaps by infusing eye-tracking data into Code LLM training.

\paragraph{Human-Centered AI for Software Engineering.}
Human input aids SE tasks (e.g., developer annotations~\cite{hamza2024human,huang2023reduce,huang2025enable}, or crowdsourcing~\cite{lu2024computing}). Eye tracking reveals detailed reading strategies~\cite{karas2024tale,li2024machines,tang2024developer,bansal2023towards}, yet most work applies these signals post hoc~\cite{bansal2023modeling} or in narrow domains~\cite{zhang2024eyetrans}. We directly integrate eye-tracking insights into LLM fine-tuning to capture genuine developer scanpaths.

\paragraph{Neural Code Analysis and Comprehension.}
Deep learning powers code summarization, clone detection, and bug identification~\cite{shi2022evaluation,richter2022neural,zhang2022pre,bansal2023revisiting,li2024malmixer}, often using tokens, ASTs, or graphs~\cite{zhang2022astro,wang2021syncobert,pailoor2024semantic,acharya2025optimizing}. These methods typically ignore real developer focus and rely on static artifacts. Our approach augments models with human attention, adding a more developer-centric lens to code comprehension.

\paragraph{LLMs for Code.}
Models such as CodeBERT~\cite{feng2020codebert}, Codex~\cite{chen2021evaluating}, and CodeT5~\cite{wang2021codet5,wang2023codet5+} excel via large-scale pretraining and selective fine-tuning but often miss actual programmer fixations. Although some studies consider integrating human attention~\cite{lai2020understanding,abulaish2024role,li2024machines}, efforts remain limited. Our reward-guided framework systematically aligns the model with genuine developer

\section{Conclusion}\label{sec:conclusion}

We present a developer-attention–driven approach for fine-tuning Code LLMs, aligning model outputs with real programmer fixations using eye-tracking data. By augmenting these signals with semantic labels and positional cues and applying reward-based fine-tuning to CodeT5, we achieve notable gains in textual and code-specific metrics. Our results underscore the potential of integrating human insight with AI, fostering deeper collaboration in AI4SE.

\bibliographystyle{ACM-Reference-Format}
\bibliography{acmart}


\begin{thebibliography}{34}

%%% the Web address.

\ifx \showCODEN    \undefined \def \showCODEN     #1{\unskip}     \fi
\ifx \showDOI      \undefined \def \showDOI       #1{#1}\fi
\ifx \showISBNx    \undefined \def \showISBNx     #1{\unskip}     \fi
\ifx \showISBNxiii \undefined \def \showISBNxiii  #1{\unskip}     \fi
\ifx \showISSN     \undefined \def \showISSN      #1{\unskip}     \fi
\ifx \showLCCN     \undefined \def \showLCCN      #1{\unskip}     \fi
\ifx \shownote     \undefined \def \shownote      #1{#1}          \fi
\ifx \showarticletitle \undefined \def \showarticletitle #1{#1}   \fi
\ifx \showURL      \undefined \def \showURL       {\relax}        \fi
% invisible to TeX
\providecommand\bibfield[2]{#2}
\providecommand\bibinfo[2]{#2}
\providecommand\natexlab[1]{#1}
\providecommand\showeprint[2][]{arXiv:#2}

\bibitem[Abulaish et~al\mbox{.}(2024)]%
        {abulaish2024role}
\bibfield{author}{\bibinfo{person}{Muhammad Abulaish}, \bibinfo{person}{Nesar~Ahmad Wasi}, {and} \bibinfo{person}{Shachi Sharma}.} \bibinfo{year}{2024}\natexlab{}.
\newblock \showarticletitle{The role of lifelong machine learning in bridging the gap between human and machine learning: A scientometric analysis}.
\newblock \bibinfo{journal}{\emph{Wiley Interdisciplinary Reviews: Data Mining and Knowledge Discovery}} \bibinfo{volume}{14}, \bibinfo{number}{2} (\bibinfo{year}{2024}), \bibinfo{pages}{e1526}.
\newblock


\bibitem[Acharya et~al\mbox{.}(2025)]%
        {acharya2025optimizing}
\bibfield{author}{\bibinfo{person}{Manish Acharya}, \bibinfo{person}{Yifan Zhang}, \bibinfo{person}{Yu Huang}, {and} \bibinfo{person}{Kevin Leach}.} \bibinfo{year}{2025}\natexlab{}.
\newblock \showarticletitle{Optimizing Code Runtime Performance through Context-Aware Retrieval-Augmented Generation}.
\newblock \bibinfo{journal}{\emph{arXiv preprint arXiv:2501.16692}} (\bibinfo{year}{2025}).
\newblock


\bibitem[Allamanis et~al\mbox{.}(2018)]%
        {allamanis2018survey}
\bibfield{author}{\bibinfo{person}{Miltiadis Allamanis}, \bibinfo{person}{Earl~T Barr}, \bibinfo{person}{Premkumar Devanbu}, {and} \bibinfo{person}{Charles Sutton}.} \bibinfo{year}{2018}\natexlab{}.
\newblock \showarticletitle{A survey of machine learning for big code and naturalness}.
\newblock \bibinfo{journal}{\emph{ACM Computing Surveys (CSUR)}} \bibinfo{volume}{51}, \bibinfo{number}{4} (\bibinfo{year}{2018}), \bibinfo{pages}{1--37}.
\newblock


\bibitem[Bansal et~al\mbox{.}(2023a)]%
        {bansal2023towards}
\bibfield{author}{\bibinfo{person}{Aakash Bansal}, \bibinfo{person}{Bonita Sharif}, {and} \bibinfo{person}{Collin McMillan}.} \bibinfo{year}{2023}\natexlab{a}.
\newblock \showarticletitle{Towards modeling human attention from eye movements for neural source code summarization}.
\newblock \bibinfo{journal}{\emph{Proceedings of the ACM on Human-Computer Interaction}} \bibinfo{volume}{7}, \bibinfo{number}{ETRA} (\bibinfo{year}{2023}), \bibinfo{pages}{1--19}.
\newblock


\bibitem[Bansal et~al\mbox{.}(2023c)]%
        {bansal2023modeling}
\bibfield{author}{\bibinfo{person}{Aakash Bansal}, \bibinfo{person}{Chia-Yi Su}, \bibinfo{person}{Zachary Karas}, \bibinfo{person}{Yifan Zhang}, \bibinfo{person}{Yu Huang}, \bibinfo{person}{Toby Jia-Jun Li}, {and} \bibinfo{person}{Collin McMillan}.} \bibinfo{year}{2023}\natexlab{c}.
\newblock \showarticletitle{Modeling programmer attention as scanpath prediction}. In \bibinfo{booktitle}{\emph{2023 38th IEEE/ACM International Conference on Automated Software Engineering (ASE)}}. IEEE, \bibinfo{pages}{1732--1736}.
\newblock


\bibitem[Bansal et~al\mbox{.}(2023b)]%
        {bansal2023revisiting}
\bibfield{author}{\bibinfo{person}{Aakash Bansal}, \bibinfo{person}{Chia-Yi Su}, {and} \bibinfo{person}{Collin McMillan}.} \bibinfo{year}{2023}\natexlab{b}.
\newblock \showarticletitle{Revisiting File Context for Source Code Summarization}.
\newblock \bibinfo{journal}{\emph{arXiv preprint arXiv:2309.02326}} (\bibinfo{year}{2023}).
\newblock


\bibitem[Chen et~al\mbox{.}(2021)]%
        {chen2021evaluating}
\bibfield{author}{\bibinfo{person}{Mark Chen}, \bibinfo{person}{Jerry Tworek}, \bibinfo{person}{Heewoo Jun}, \bibinfo{person}{Qiming Yuan}, \bibinfo{person}{Henrique Ponde De~Oliveira Pinto}, \bibinfo{person}{Jared Kaplan}, \bibinfo{person}{Harri Edwards}, \bibinfo{person}{Yuri Burda}, \bibinfo{person}{Nicholas Joseph}, \bibinfo{person}{Greg Brockman}, {et~al\mbox{.}}} \bibinfo{year}{2021}\natexlab{}.
\newblock \showarticletitle{Evaluating large language models trained on code}.
\newblock \bibinfo{journal}{\emph{arXiv preprint arXiv:2107.03374}} (\bibinfo{year}{2021}).
\newblock


\bibitem[Feng et~al\mbox{.}(2020)]%
        {feng2020codebert}
\bibfield{author}{\bibinfo{person}{Zhangyin Feng}, \bibinfo{person}{Daya Guo}, \bibinfo{person}{Duyu Tang}, \bibinfo{person}{Nan Duan}, \bibinfo{person}{Xiaocheng Feng}, \bibinfo{person}{Ming Gong}, \bibinfo{person}{Linjun Shou}, \bibinfo{person}{Bing Qin}, \bibinfo{person}{Ting Liu}, \bibinfo{person}{Daxin Jiang}, {et~al\mbox{.}}} \bibinfo{year}{2020}\natexlab{}.
\newblock \showarticletitle{Codebert: A pre-trained model for programming and natural languages}.
\newblock \bibinfo{journal}{\emph{arXiv preprint arXiv:2002.08155}} (\bibinfo{year}{2020}).
\newblock


\bibitem[Hamza et~al\mbox{.}(2024)]%
        {hamza2024human}
\bibfield{author}{\bibinfo{person}{Muhammad Hamza}, \bibinfo{person}{Dominik Siemon}, \bibinfo{person}{Muhammad~Azeem Akbar}, {and} \bibinfo{person}{Tahsinur Rahman}.} \bibinfo{year}{2024}\natexlab{}.
\newblock \showarticletitle{Human-ai collaboration in software engineering: Lessons learned from a hands-on workshop}. In \bibinfo{booktitle}{\emph{Proceedings of the 7th ACM/IEEE International Workshop on Software-intensive Business}}. \bibinfo{pages}{7--14}.
\newblock


\bibitem[He et~al\mbox{.}(2024)]%
        {he2024exploring}
\bibfield{author}{\bibinfo{person}{Pengfei He}, \bibinfo{person}{Shaowei Wang}, \bibinfo{person}{Shaiful Chowdhury}, {and} \bibinfo{person}{Tse-Hsun Chen}.} \bibinfo{year}{2024}\natexlab{}.
\newblock \showarticletitle{Exploring Demonstration Retrievers in RAG for Coding Tasks: Yeas and Nays!}
\newblock \bibinfo{journal}{\emph{arXiv preprint arXiv:2410.09662}} (\bibinfo{year}{2024}).
\newblock


\bibitem[Hoq et~al\mbox{.}(2024)]%
        {hoq2024towards}
\bibfield{author}{\bibinfo{person}{Muntasir Hoq}, \bibinfo{person}{Jessica Vandenberg}, \bibinfo{person}{Bradford Mott}, \bibinfo{person}{James Lester}, \bibinfo{person}{Narges Norouzi}, {and} \bibinfo{person}{Bita Akram}.} \bibinfo{year}{2024}\natexlab{}.
\newblock \showarticletitle{Towards Attention-Based Automatic Misconception Identification in Introductory Programming Courses}. In \bibinfo{booktitle}{\emph{Proceedings of the 55th ACM Technical Symposium on Computer Science Education V. 2}}. \bibinfo{pages}{1680--1681}.
\newblock


\bibitem[Huang et~al\mbox{.}(2025)]%
        {huang2025enable}
\bibfield{author}{\bibinfo{person}{Chen Huang}, \bibinfo{person}{Yang Deng}, \bibinfo{person}{Wenqiang Lei}, \bibinfo{person}{Jiancheng Lv}, \bibinfo{person}{Tat-Seng Chua}, {and} \bibinfo{person}{Jimmy~Xiangji Huang}.} \bibinfo{year}{2025}\natexlab{}.
\newblock \showarticletitle{How to Enable Effective Cooperation Between Humans and NLP Models: A Survey of Principles, Formalizations, and Beyond}.
\newblock \bibinfo{journal}{\emph{arXiv preprint arXiv:2501.05714}} (\bibinfo{year}{2025}).
\newblock


\bibitem[Huang et~al\mbox{.}(2023)]%
        {huang2023reduce}
\bibfield{author}{\bibinfo{person}{Chen Huang}, \bibinfo{person}{Peixin Qin}, \bibinfo{person}{Wenqiang Lei}, {and} \bibinfo{person}{Jiancheng Lv}.} \bibinfo{year}{2023}\natexlab{}.
\newblock \showarticletitle{Reduce Human Labor On Evaluating Conversational Information Retrieval System: A Human-Machine Collaboration Approach}. In \bibinfo{booktitle}{\emph{Proceedings of the 2023 Conference on Empirical Methods in Natural Language Processing}}. \bibinfo{pages}{10876--10891}.
\newblock


\bibitem[Karas et~al\mbox{.}(2024)]%
        {karas2024tale}
\bibfield{author}{\bibinfo{person}{Zachary Karas}, \bibinfo{person}{Aakash Bansal}, \bibinfo{person}{Yifan Zhang}, \bibinfo{person}{Toby Li}, \bibinfo{person}{Collin McMillan}, {and} \bibinfo{person}{Yu Huang}.} \bibinfo{year}{2024}\natexlab{}.
\newblock \showarticletitle{A Tale of Two Comprehensions? Analyzing Student Programmer Attention during Code Summarization}.
\newblock \bibinfo{journal}{\emph{ACM Transactions on Software Engineering and Methodology}} (\bibinfo{year}{2024}).
\newblock


\bibitem[Kou et~al\mbox{.}(2024)]%
        {kou2024large}
\bibfield{author}{\bibinfo{person}{Bonan Kou}, \bibinfo{person}{Shengmai Chen}, \bibinfo{person}{Zhijie Wang}, \bibinfo{person}{Lei Ma}, {and} \bibinfo{person}{Tianyi Zhang}.} \bibinfo{year}{2024}\natexlab{}.
\newblock \showarticletitle{Do large language models pay similar attention like human programmers when generating code?}
\newblock \bibinfo{journal}{\emph{Proceedings of the ACM on Software Engineering}} \bibinfo{volume}{1}, \bibinfo{number}{FSE} (\bibinfo{year}{2024}), \bibinfo{pages}{2261--2284}.
\newblock


\bibitem[Lai et~al\mbox{.}(2020)]%
        {lai2020understanding}
\bibfield{author}{\bibinfo{person}{Qiuxia Lai}, \bibinfo{person}{Salman Khan}, \bibinfo{person}{Yongwei Nie}, \bibinfo{person}{Hanqiu Sun}, \bibinfo{person}{Jianbing Shen}, {and} \bibinfo{person}{Ling Shao}.} \bibinfo{year}{2020}\natexlab{}.
\newblock \showarticletitle{Understanding more about human and machine attention in deep neural networks}.
\newblock \bibinfo{journal}{\emph{IEEE Transactions on Multimedia}}  \bibinfo{volume}{23} (\bibinfo{year}{2020}), \bibinfo{pages}{2086--2099}.
\newblock


\bibitem[Li et~al\mbox{.}(2024a)]%
        {li2024malmixer}
\bibfield{author}{\bibinfo{person}{Jiliang Li}, \bibinfo{person}{Yifan Zhang}, \bibinfo{person}{Yu Huang}, {and} \bibinfo{person}{Kevin Leach}.} \bibinfo{year}{2024}\natexlab{a}.
\newblock \showarticletitle{Malmixer: Few-shot malware classification with retrieval-augmented semi-supervised learning}.
\newblock \bibinfo{journal}{\emph{arXiv preprint arXiv:2409.13213}} (\bibinfo{year}{2024}).
\newblock


\bibitem[Li et~al\mbox{.}(2024b)]%
        {li2024machines}
\bibfield{author}{\bibinfo{person}{Jiliang Li}, \bibinfo{person}{Yifan Zhang}, \bibinfo{person}{Zachary Karas}, \bibinfo{person}{Collin McMillan}, \bibinfo{person}{Kevin Leach}, {and} \bibinfo{person}{Yu Huang}.} \bibinfo{year}{2024}\natexlab{b}.
\newblock \showarticletitle{Do Machines and Humans Focus on Similar Code? Exploring Explainability of Large Language Models in Code Summarization}. In \bibinfo{booktitle}{\emph{Proceedings of the 32nd IEEE/ACM International Conference on Program Comprehension}}. \bibinfo{pages}{47--51}.
\newblock


\bibitem[Lu et~al\mbox{.}(2024)]%
        {lu2024computing}
\bibfield{author}{\bibinfo{person}{Yao Lu}, \bibinfo{person}{Song Bian}, \bibinfo{person}{Lequn Chen}, \bibinfo{person}{Yongjun He}, \bibinfo{person}{Yulong Hui}, \bibinfo{person}{Matthew Lentz}, \bibinfo{person}{Beibin Li}, \bibinfo{person}{Fei Liu}, \bibinfo{person}{Jialin Li}, \bibinfo{person}{Qi Liu}, {et~al\mbox{.}}} \bibinfo{year}{2024}\natexlab{}.
\newblock \showarticletitle{Computing in the Era of Large Generative Models: From Cloud-Native to AI-Native}.
\newblock \bibinfo{journal}{\emph{arXiv preprint arXiv:2401.12230}} (\bibinfo{year}{2024}).
\newblock


\bibitem[Minelli et~al\mbox{.}(2015)]%
        {minelli2015know}
\bibfield{author}{\bibinfo{person}{Roberto Minelli}, \bibinfo{person}{Andrea Mocci}, {and} \bibinfo{person}{Michele Lanza}.} \bibinfo{year}{2015}\natexlab{}.
\newblock \showarticletitle{I know what you did last summer-an investigation of how developers spend their time}. In \bibinfo{booktitle}{\emph{2015 IEEE 23rd international conference on program comprehension}}. IEEE, \bibinfo{pages}{25--35}.
\newblock


\bibitem[Nijkamp et~al\mbox{.}(2023)]%
        {nijkamp2023codegen2}
\bibfield{author}{\bibinfo{person}{Erik Nijkamp}, \bibinfo{person}{Hiroaki Hayashi}, \bibinfo{person}{Caiming Xiong}, \bibinfo{person}{Silvio Savarese}, {and} \bibinfo{person}{Yingbo Zhou}.} \bibinfo{year}{2023}\natexlab{}.
\newblock \showarticletitle{Codegen2: Lessons for training llms on programming and natural languages}.
\newblock \bibinfo{journal}{\emph{arXiv preprint arXiv:2305.02309}} (\bibinfo{year}{2023}).
\newblock


\bibitem[Pailoor et~al\mbox{.}(2024)]%
        {pailoor2024semantic}
\bibfield{author}{\bibinfo{person}{Shankara Pailoor}, \bibinfo{person}{Yuepeng Wang}, {and} \bibinfo{person}{I{\c{s}}{\i}l Dillig}.} \bibinfo{year}{2024}\natexlab{}.
\newblock \showarticletitle{Semantic code refactoring for abstract data types}.
\newblock \bibinfo{journal}{\emph{Proceedings of the ACM on Programming Languages}} \bibinfo{volume}{8}, \bibinfo{number}{POPL} (\bibinfo{year}{2024}), \bibinfo{pages}{816--847}.
\newblock


\bibitem[Ren et~al\mbox{.}(2020)]%
        {ren2020codebleu}
\bibfield{author}{\bibinfo{person}{Shuo Ren}, \bibinfo{person}{Daya Guo}, \bibinfo{person}{Shuai Lu}, \bibinfo{person}{Long Zhou}, \bibinfo{person}{Shujie Liu}, \bibinfo{person}{Duyu Tang}, \bibinfo{person}{Neel Sundaresan}, \bibinfo{person}{Ming Zhou}, \bibinfo{person}{Ambrosio Blanco}, {and} \bibinfo{person}{Shuai Ma}.} \bibinfo{year}{2020}\natexlab{}.
\newblock \showarticletitle{Codebleu: a method for automatic evaluation of code synthesis}.
\newblock \bibinfo{journal}{\emph{arXiv preprint arXiv:2009.10297}} (\bibinfo{year}{2020}).
\newblock


\bibitem[Richter et~al\mbox{.}(2022)]%
        {richter2022neural}
\bibfield{author}{\bibinfo{person}{Cedric Richter}, \bibinfo{person}{Jan Haltermann}, \bibinfo{person}{Marie-Christine Jakobs}, \bibinfo{person}{Felix Pauck}, \bibinfo{person}{Stefan Schott}, {and} \bibinfo{person}{Heike Wehrheim}.} \bibinfo{year}{2022}\natexlab{}.
\newblock \showarticletitle{Are Neural Bug Detectors Comparable to Software Developers on Variable Misuse Bugs?}. In \bibinfo{booktitle}{\emph{Proceedings of the 37th IEEE/ACM International Conference on Automated Software Engineering}}. \bibinfo{pages}{1--12}.
\newblock


\bibitem[Salvucci and Goldberg(2000)]%
        {salvucci2000identifying}
\bibfield{author}{\bibinfo{person}{Dario~D Salvucci} {and} \bibinfo{person}{Joseph~H Goldberg}.} \bibinfo{year}{2000}\natexlab{}.
\newblock \showarticletitle{Identifying fixations and saccades in eye-tracking protocols}. In \bibinfo{booktitle}{\emph{Proceedings of the 2000 symposium on Eye tracking research \& applications}}. \bibinfo{pages}{71--78}.
\newblock


\bibitem[Shi et~al\mbox{.}(2022)]%
        {shi2022evaluation}
\bibfield{author}{\bibinfo{person}{Ensheng Shi}, \bibinfo{person}{Yanlin Wang}, \bibinfo{person}{Lun Du}, \bibinfo{person}{Junjie Chen}, \bibinfo{person}{Shi Han}, \bibinfo{person}{Hongyu Zhang}, \bibinfo{person}{Dongmei Zhang}, {and} \bibinfo{person}{Hongbin Sun}.} \bibinfo{year}{2022}\natexlab{}.
\newblock \showarticletitle{On the evaluation of neural code summarization}. In \bibinfo{booktitle}{\emph{Proceedings of the 44th international conference on software engineering}}. \bibinfo{pages}{1597--1608}.
\newblock


\bibitem[Tang et~al\mbox{.}(2024)]%
        {tang2024developer}
\bibfield{author}{\bibinfo{person}{Ningzhi Tang}, \bibinfo{person}{Meng Chen}, \bibinfo{person}{Zheng Ning}, \bibinfo{person}{Aakash Bansal}, \bibinfo{person}{Yu Huang}, \bibinfo{person}{Collin McMillan}, {and} \bibinfo{person}{Toby Jia-Jun Li}.} \bibinfo{year}{2024}\natexlab{}.
\newblock \showarticletitle{Developer behaviors in validating and repairing llm-generated code using ide and eye tracking}. In \bibinfo{booktitle}{\emph{2024 IEEE Symposium on Visual Languages and Human-Centric Computing (VL/HCC)}}. IEEE, \bibinfo{pages}{40--46}.
\newblock


\bibitem[Wang et~al\mbox{.}(2021b)]%
        {wang2021syncobert}
\bibfield{author}{\bibinfo{person}{Xin Wang}, \bibinfo{person}{Yasheng Wang}, \bibinfo{person}{Fei Mi}, \bibinfo{person}{Pingyi Zhou}, \bibinfo{person}{Yao Wan}, \bibinfo{person}{Xiao Liu}, \bibinfo{person}{Li Li}, \bibinfo{person}{Hao Wu}, \bibinfo{person}{Jin Liu}, {and} \bibinfo{person}{Xin Jiang}.} \bibinfo{year}{2021}\natexlab{b}.
\newblock \showarticletitle{Syncobert: Syntax-guided multi-modal contrastive pre-training for code representation}.
\newblock \bibinfo{journal}{\emph{arXiv preprint arXiv:2108.04556}} (\bibinfo{year}{2021}).
\newblock


\bibitem[Wang et~al\mbox{.}(2023)]%
        {wang2023codet5+}
\bibfield{author}{\bibinfo{person}{Yue Wang}, \bibinfo{person}{Hung Le}, \bibinfo{person}{Akhilesh~Deepak Gotmare}, \bibinfo{person}{Nghi~DQ Bui}, \bibinfo{person}{Junnan Li}, {and} \bibinfo{person}{Steven~CH Hoi}.} \bibinfo{year}{2023}\natexlab{}.
\newblock \showarticletitle{Codet5+: Open code large language models for code understanding and generation}.
\newblock \bibinfo{journal}{\emph{arXiv preprint arXiv:2305.07922}} (\bibinfo{year}{2023}).
\newblock


\bibitem[Wang et~al\mbox{.}(2021a)]%
        {wang2021codet5}
\bibfield{author}{\bibinfo{person}{Yue Wang}, \bibinfo{person}{Weishi Wang}, \bibinfo{person}{Shafiq Joty}, {and} \bibinfo{person}{Steven~CH Hoi}.} \bibinfo{year}{2021}\natexlab{a}.
\newblock \showarticletitle{Codet5: Identifier-aware unified pre-trained encoder-decoder models for code understanding and generation}.
\newblock \bibinfo{journal}{\emph{arXiv preprint arXiv:2109.00859}} (\bibinfo{year}{2021}).
\newblock


\bibitem[Yu et~al\mbox{.}(2022)]%
        {yu2022data}
\bibfield{author}{\bibinfo{person}{Shiwen Yu}, \bibinfo{person}{Ting Wang}, {and} \bibinfo{person}{Ji Wang}.} \bibinfo{year}{2022}\natexlab{}.
\newblock \showarticletitle{Data augmentation by program transformation}.
\newblock \bibinfo{journal}{\emph{Journal of Systems and Software}}  \bibinfo{volume}{190} (\bibinfo{year}{2022}), \bibinfo{pages}{111304}.
\newblock


\bibitem[Zhang et~al\mbox{.}(2022a)]%
        {zhang2022pre}
\bibfield{author}{\bibinfo{person}{Yifan Zhang}, \bibinfo{person}{Chen Huang}, \bibinfo{person}{Kevin Cao}, \bibinfo{person}{Yueke Zhang}, \bibinfo{person}{Scott~Thomas Andersen}, \bibinfo{person}{Huajie Shao}, \bibinfo{person}{Kevin Leach}, {and} \bibinfo{person}{Yu Huang}.} \bibinfo{year}{2022}\natexlab{a}.
\newblock \showarticletitle{Pre-Training Representations of Binary Code Using Contrastive Learning}.
\newblock \bibinfo{journal}{\emph{arXiv preprint arXiv:2210.05102}} (\bibinfo{year}{2022}).
\newblock


\bibitem[Zhang et~al\mbox{.}(2024)]%
        {zhang2024eyetrans}
\bibfield{author}{\bibinfo{person}{Yifan Zhang}, \bibinfo{person}{Jiliang Li}, \bibinfo{person}{Zachary Karas}, \bibinfo{person}{Aakash Bansal}, \bibinfo{person}{Toby Jia-Jun Li}, \bibinfo{person}{Collin McMillan}, \bibinfo{person}{Kevin Leach}, {and} \bibinfo{person}{Yu Huang}.} \bibinfo{year}{2024}\natexlab{}.
\newblock \showarticletitle{Eyetrans: Merging human and machine attention for neural code summarization}.
\newblock \bibinfo{journal}{\emph{Proceedings of the ACM on Software Engineering}} \bibinfo{volume}{1}, \bibinfo{number}{FSE} (\bibinfo{year}{2024}), \bibinfo{pages}{115--136}.
\newblock


\bibitem[Zhang et~al\mbox{.}(2022b)]%
        {zhang2022astro}
\bibfield{author}{\bibinfo{person}{Yifan Zhang}, \bibinfo{person}{Junwen Yang}, \bibinfo{person}{Haoyu Dong}, \bibinfo{person}{Qingchen Wang}, \bibinfo{person}{Huajie Shao}, \bibinfo{person}{Kevin Leach}, {and} \bibinfo{person}{Yu Huang}.} \bibinfo{year}{2022}\natexlab{b}.
\newblock \showarticletitle{Astro: An ast-assisted approach for generalizable neural clone detection}.
\newblock \bibinfo{journal}{\emph{arXiv preprint arXiv:2208.08067}} (\bibinfo{year}{2022}).
\newblock


\end{thebibliography}

\end{document}